\documentclass[aps,prd,reprint,nofootinbib,bibnotes,groupedaddress]{revtex4-2}
\usepackage{amsmath,bbm,hyperref,mathrsfs,graphicx}
\begin{document}

\title{Isospectral local Hermitian theory for the $\mathcal{PT}$-symmetric $i\phi^3$ quantum field theory}

\author{Yi-Da Li}
\email{liyd20@mails.tsinghua.edu.cn}
\affiliation{Department of Physics, Tsinghua University, Beijing 100084, P. R. China}

\author{Qing Wang}
\email[Corresponding author:~]{wangq@mail.tsinghua.edu.cn}
\affiliation{Department of Physics, Tsinghua University, Beijing 100084, P. R. China \\
Center for High Energy Physics, Tsinghua University, Beijing 100084, P. R. China}

\date{\today}

\begin{abstract}
We propose a new method to calculate perturbatively the isospectral Hermitian theory for the $\mathcal{PT}$-symmetric $i\phi^3$ quantum field theory in $d$ dimensions, whose result is local. The result of the new method in $1$ dimension reproduces our previous result in the $ix^3$ quantum mechanics, and the new method can be seen as a generalization of our previous method to quantum field theory. We also find the isospectral local Hermitian theory has the same form in all dimensions and differs in coefficients only, and our previous results in quantum mechanics can be used directly to determine the form of the isospectral local Hermitian quantum field theory.
\end{abstract}

\maketitle
\section{Introduction\label{sec-intro}}
The study of $\mathcal{PT}$-symmetric quantum theory originates from Bender and Boettcher's groundbreaking work\cite{bender1998}, where the real and positive spectrum of $H=p^2-(ix)^N,\ (N\ge2),$ was attributed to the Hamiltonian's $\mathcal{PT}$ symmetry. It was soon realized that $\mathcal{PT}$-symmetric quantum theory is a new kind of physical theory apart from Hermitian quantum theory, and the general framework describing $\mathcal{PT}$-symmetric quantum theory was established\cite{ali2002a,ali2002b,ali2002c,ali2003a,ali2003b,ali2004,bender2002,bender2004a,bender2004b,bender2005}. A $\mathcal{PT}$-symmetric Hamiltonian $H$ is also pseudo-Hermitian\cite{ali2002c} satisfying $VHV^{-1}=H^\dag$ for an invertible operator $V$. If the $\mathcal{PT}$ symmetry of $H$ is unbroken, $V$ is Hermitian and can be decomposed into $V=\eta^\dag\eta$\cite{ali2003a} by an invertible operator $\eta$, and we can define a Hermitian Hamiltonian $h=\eta H\eta^{-1}$ which is isospectral to $H$. Therefore, seeking for the isospectral Hermitian Hamiltonian for a $\mathcal{PT}$-symmetric Hamiltonian is a key procedure in studying $\mathcal{PT}$-symmetric quantum theory. A significant result is the isospectral Hermitian Hamiltonian of $H=\frac{p^2}{2m}-gx^4$, which is found to be $h=\frac{\tilde{p}^2}{2m}-\hbar\sqrt{\frac{2g}{m}}z+4gz^4$\cite{bender2006,jones2006a,jones2006b}. The equivalence of the $\mathcal{PT}$-symmetric $-x^4$ and the Hermitian $x^4$ inspired the solution of standard model vacuum stability problem caused by the $-\phi^4$ potential of the Higgs particle in high energy\cite{bender-book}. However, nonperturbative methods must be used for the $\mathcal{PT}$-symmetric $-\phi^4$ theory\cite{bender-book}, which renders the study of the $\mathcal{PT}$-symmetric $-\phi^4$  theory rather difficult. The study of the $\mathcal{PT}$-symmetric $-\phi^4$ theory is still ongoing\cite{shalaby2009a,shalaby2010,shalaby2013,bender2018,bender2021,bender2022,paul2023a,paul2023b}.

In perturbative $\mathcal{PT}$-symmetric quantum theories, the calculation of isospectral Hermitian Hamiltonians is simpler and is already formulated in a systematic way\cite{bender2004b,bender-book}. This traditional method focuses on $\mathcal{PT}$-symmetric Hamiltonians of the form $H=H_0+gH_1$, where $H_0$ is the Hermitian free Hamiltonian, $H_1$ is the anti-Hermitian interaction Hamiltonian and $g$ is the real coupling constant. Set $V=e^{-Q}$ and expand $Q$ perturbatively as $Q=\sum_{k=0}^\infty g^{2k+1}Q_{2k+1}$, and $Q_{2k+1}$  can be solved from the pseudo-Hermiticity of $H$ resulting equations as follows\cite{bender2004b}:
\begin{equation}
[H_0,Q_1]=-2H_1,[H_0,Q_3]=-\frac16[[H_1,Q_1],Q_1],\cdots.
\end{equation}
Then we have $h=e^{-\frac{Q}{2}}He^{\frac{Q}{2}}$ as an isospectral Hermitian Hamiltonian of $H$.

The $\mathcal{PT}$-symmetric $i\phi^3$ theory is a perturbative theory to which the traditional method can be applied, either in quantum mechanics or in quantum field theory. In recent years, the $\mathcal{PT}$-symmetric $i\phi^3$ theory is studied frequently, about its isospectral Hermitian theory\cite{bender2004a,bender2004b,bender2003,ali2005,jones2005,shalaby2009b,bender2009,siegl2012,bender2012}, critical behavior\cite{bender2013}, running behavior of the coupling constant\cite{shalaby2020,aditya2021}, effective potential\cite{shalaby2017,shalaby2019}, and scattering amplitudes\cite{novikov2019}. However, the isospectral Hermitian theory of the $\mathcal{PT}$-symmetric $i\phi^3$ theory from the traditional method, has unusual forms. In the $\mathcal{PT}$-symmetric $ix^3$ quantum mechanics, the isospectral Hermitian Hamiltonian is complicated by the appearance of momentum $p$ in the potential\cite{ali2005,bender-book}. In the $\mathcal{PT}$-symmetric $i\phi^3$ quantum field theory, the equivalent Hermitian Hamiltonian is nonlocal\cite{bender2004a,bender2004b,bender-book}, apart from the appearance of field momentum. For convenience, we call the appearance of momentum as nonlocality, too. These nonlocal forms complicate the physical meaning of the isospectral Hermitian theory. 

In our previous work\cite{lyd2023}, we developed a method to work out isospectral local Hermitian Hamiltonians of perturbative $\mathcal{PT}$-symmetric Hamiltonians in quantum mechanics. In this work, we generalize our method to perturbative $\mathcal{PT}$-symmetric quantum field theory and work out the isospectral local Hermitian theory of the $\mathcal{PT}$-symmetric $i\phi^3$ theory up to $g^2$-order. In our previous work, we transform a $\mathcal{PT}$-symmetric Hamiltonian $H_V=\frac12p^2+\frac12m^2x^2+\sum_{n=1}^\infty g^nV_n(x,p)$ into a diagonal Hamiltonian $H_N=m(N+\frac12)+\sum_{n=1}^\infty g^nf_n(N)$ where $N$ is the number operator in the Fock representation, and then we transform the diagonal Hamiltonian into a local Hermitian Hamiltonian $h_v=\frac12p^2+\frac12m^2x^2+\sum_{n=1}^\infty g^nv_n(x)$, making use of a one-to-one correspondence between polynomials of $N$ and polynomials of $x^2$. The diagonal form of the $\mathcal{PT}$-symmetric Hamiltonian, is a vital intermediate form in the whole procedure. Designing the isospectral local Hermitian Hamiltonian of a $\mathcal{PT}$-symmetric Hamiltonian directly may be hard, but diagonalizing a Hamiltonian is easy. If a $\mathcal{PT}$-symmetric Hamiltonian and a local Hermitian Hamiltonian both have the same diagonalized form, they must be related by a similarity transformation and we find the isospectral pairs. However, as pointed out in our previous work\cite{lyd2023}, our method in the Hamiltonian form failed to generalize to multi-variable quantum mechanics with degeneracy in the free Hamiltonian's energy spectrum, and consequently our method cannot be used directly in quantum field theory where there are many degenerate energy eigenstates caused by Lorentz symmetry. Nevertheless, we find that degenerate energy eigenstates are not problematic in the path integral formalism where our method is still applicable. The main idea is the same as that in quantum mechanics: if a $\mathcal{PT}$-symmetric quantum field theory and a local Hermitian quantum field theory both can be transformed into the same free form, they must be related by a valid transformation and we find the isospectral pairs.

This paper is organized as follows. In Sec. \ref{sec-gene} we describe the transformation that transforms the $\mathcal{PT}$-symmetric $i\phi^3$ theory into its isospectral local Hermitian theory in $d$ dimensions. In Sec. \ref{sec-d0} we work in $0$ dimensions and solve the transformation by numerical methods. In Sec. \ref{sec-sum} we conclude.

\begin{widetext}
\section{General analysis in $d$ dimensions}\label{sec-gene}
We start with the Euclidean partition function of the $i\phi^3$ in $d$ dimensions,
\begin{equation}\label{eq-ziphi3}\begin{aligned}
Z_{\mathcal{PT}}[j]=\int D\phi\exp\left\{-\int d^dx\left[\frac12\phi_x\left(-\partial^2_x+m^2\right)\phi_x+ig\phi^3_x+i\phi_{\mathrm{phys},x}j_x\right]\right\},
\end{aligned}\end{equation}
where $\phi_x\equiv\phi(x)$ is the real pseudoscalar field, $j_x\equiv j(x)$ is the external source, $\phi_{\mathrm{phys},x}\equiv\phi_{\mathrm{phys}}(x)$ is the physical field which is related to $\phi_x$ by $\phi_{\mathrm{phys},x}=\eta\phi_x\eta^{-1}$ in the operator formalism. It is the coupling of $j_x$ to $\phi_{\mathrm{phys},x}$ rather than that to $\phi_x$ that leads to physical observables, which has already been demonstrated\cite{jones2007,ali2007}. We assume that $\phi_{\mathrm{phys},x}=\eta(\phi_x)$, which will be solved later.

Similar to our method\cite{lyd2023} in quantum mechanics, we first transform $Z_{\mathcal{PT}}[j]$ into a form with only local quadratic terms except for terms coupling to $j_x$, and then seek for the isospectral local Hermitian counterpart. 

We are inspired from the simpler $i\phi$ theory. Consider the following Eucliden partition function
\begin{equation}\begin{aligned}
Z_{\mathcal{PT},i\phi}[j]=\int D\phi\exp\left\{-\int d^dx\left[\frac12\phi_x\left(-\partial^2_x+m^2\right)\phi_x+ig\phi_x+i\phi_{\mathrm{phys},x}j_x\right]\right\},
\end{aligned}\end{equation}
which is easily transformed into a purely quadratic form as follows,
\begin{equation}\label{eq-iphi}\begin{aligned}
Z_{\mathcal{PT},i\phi}[j]=&\int D\phi\exp\left\{-\int d^dx\left[\frac12\left(\phi_x+\frac{ig}{m^2}\right)\left(-\partial^2_x+m^2\right)\left(\phi_x+\frac{ig}{m^2}\right)+\frac{g^2}{2m^2}+i\eta_{i\phi}(\phi_x)j_x\right]\right\}\\
=&\int D\phi\det\left(\frac{\delta\left(\phi_x-\frac{ig}{m^2}\right)}{\delta\phi_y}\right)\exp\left\{-\int d^dx\left[\frac12\phi_x\left(-\partial^2_x+m^2\right)\phi_x+\frac{g^2}{2m^2}+i\eta_{i\phi}\left(\phi_x-\frac{ig}{m^2}\right)j_x\right]\right\},
\end{aligned}\end{equation}
where $\phi_{\mathrm{phys},x}=\eta_{i\phi}(\phi_x)=\phi_x+\frac{ig}{m^2}$ can be easily read from \eqref{eq-iphi} by setting $\eta_{i\phi}\left(\phi_x-\frac{ig}{m^2}\right)=\phi_x$ because $\phi_x$ after transformation is already the physical field. \eqref{eq-iphi} is simply a completing-the-square procedure which should also be useful in the $i\phi^3$ theory.

Tentatively, we make a transformation $\phi_x\rightarrow\phi_x-ig\int d^dyD_{x-y}\phi^2_y$ in \eqref{eq-ziphi3}, where $D_{x-y}\equiv\int\frac{d^dp}{(2\pi)^d}\frac{e^{ip(x-y)}}{p^2+m^2}$ is the propagator satisfying $\left(-\partial^2_x+m^2\right)D_{x-y}=\delta^d(x-y)$, and the result is as follows,
\begin{equation}\begin{aligned}
Z_{\mathcal{PT}}[j]=\int D\phi&\det\left(\frac{\delta\left(\phi_x-ig\int d^dzD_{x-z}\phi^2_z\right)}{\delta\phi_y}\right)\exp\left\{-\int d^dx\left[\frac12\phi_x\left(-\partial^2_x+m^2\right)\phi_x\right.\right.\\
&\left.\left.+\frac52g^2\int d^dy\phi^2_xD_{x-y}\phi^2_y+i\eta\left(\phi_x-ig\int d^dyD_{x-y}\phi^2_y\right)j_x+\mathcal{O}(g^3)\right]\right\}.
\end{aligned}\end{equation}
The $ig\phi^3_x$ term is cancelled and $g^2$-order terms appear. We can absorb these $g^2$-order terms again by completinig-the-square procedure, with the introduction of higher order terms. Finally we end with a transformation which is an infinite series. In this work , we study  terms up to $g^2$ order only, and leave the study of higher order terms to future works. 

However, the measure of the path integral has a nontrivial Jacobian,
\begin{equation}\begin{aligned}
\det\left(\frac{\delta\left(\phi_x-ig\int d^dzD_{x-z}\phi^2_z\right)}{\delta\phi_y}\right)=&\det\left(\delta^d(x-y)-2igD_{x-y}\phi_y\right)\\
=&\exp\mathrm{tr}\ln\left(\delta^d(x-y)-2igD_{x-y}\phi_y\right)\\
=&\exp\left\{-\int d^dx\left[2igD_0\phi_x-2g^2\int d^dy\phi_xD^2_{x-y}\phi_y\right]\right\},
\end{aligned}\end{equation}
which also contributes to the path integral and should be compensated by additional terms in the transformation. Therefore, we make a general ansatz of the transformation $\phi_x\rightarrow\rho(\phi_x)$ which is expcted to transform the $i\phi^3$ theory into a free theory,
\begin{equation}\label{eq-transrho}\begin{aligned}
\rho\left(\phi_x\right)=&\phi_x+\int d^dy\left[c_1igD_{x-y}\phi^2_y+c_2g^2\int d^dzD_{x-y}\phi_yD_{y-z}\phi^2_z\right.\\
&\left.+c_3ig D_0D_{x-y}+c_4g^2D_0\int d^dzD_{x-y}\phi_yD_{y-z}+c_5g^2\int d^dzD_{x-y}D_{y-z}^2\phi_z+\mathcal{O}(g^3)\right],
\end{aligned}\end{equation}
where $c_1\sim c_5$ are real coefficients. If $\phi_x$ is on the real axis, $\rho(\phi_x)$ is no longer on the real axis, which brings the question that whether $\rho(\phi_x)$ is in the Stokes sector\cite{bender-book} containing the real axis or not. Moreover, the existence of $\rho(\phi_x)$ is also important, which justifies the use of perturbation theory. We will show numerically that $\rho(\phi_x)$ exists and is in the Stokes sector containing the real axis in $0$ dimensions in Sec. \ref{sec-d0}, and leave the existence problem and asymptotic behavior in higher dimensions to future works.

Under the transformation \eqref{eq-transrho}, $Z_{\mathcal{PT}}[j]$ is as follows,
\begin{equation}\label{eq-ziphi3t}\begin{aligned}
Z_{\mathcal{PT}}[j]=&\int D\phi\exp\left\{-\int d^dx\left[\frac12\phi_x\left(-\partial^2_x+m^2\right)\phi_x+\eta(\rho(\phi_x))j_x\right.\right.\\
&+(-c_2-3c_3+c_4-c_1c_3)\frac{g^2}{m^2}D_0\phi^2_x+\left(-c_4-\frac12c_3^2\right)\frac{g^2}{m^2}D_0^2-c_5g^2\int d^dyD_{y}^3\\
&+\left(1+c_1\right)igD_0\phi^3_x+\left(-2c_1+c_3\right)igD_0\phi_x\\
&\left.\left.+(-2c_1^2-2c_2+c_5)g^2\int d^dy\phi_xD_{x-y}^2\phi_y+\left(-\frac12c_1^2-3c_1+c_2\right)g^2\int d^dy\phi^2_xD_{x-y}\phi^2_y+\mathcal{O}(g^3)\right]\right\}.
\end{aligned}\end{equation}
To obtain a local Hermitian quadratic form, the following equations must be satisfied,
\begin{equation}\begin{aligned}
&1+c_1=0,\ -2c_1+c_3=0,\ -2c_1^2-2c_2+c_5=0,\ -\frac12c_1^2-3c_1+c_2=0,
\end{aligned}\end{equation}
from which we have,
\begin{equation}\begin{aligned}
c_1=-1,\ c_2=-\frac52,\ c_3=-2\ ,c_5=-3.
\end{aligned}\end{equation}
The requirement that all nonlocal terms in \eqref{eq-ziphi3t} vanish needs some explanations. There are two $g^2$-order nonlocal terms in \eqref{eq-ziphi3t}, which are $g^2\int d^dx  d^dy\phi_xD_{x-y}^2\phi_y$ and $g^2\int d^dx d^dy\phi^2_xD_{x-y}\phi^2_y$. The forms of these two terms depend entirely on the interaction term in the original action, \emph{i.e.}, $\int d^dxig\phi^3_x$ in \eqref{eq-ziphi3}. Actually, from the Feynman diagrammatic perspective, $g^2\int d^dx d^dy\phi_xD_{x-y}^2\phi_y$ is the 1-loop diagram formed by two $\int d^dxig\phi^3_x$ vertices with two uncontracted fields, and $g^2\int d^dx d^dy\phi^2_xD_{x-y}\phi^2_y$ is the tree diagram formed by two $\int d^dxig\phi^3_x$ vertices with four uncontracted fields. If we require $Z_{\mathcal{PT}}[j]$ in \eqref{eq-ziphi3t} has the most general form so that we can transform the assumed isospectral Hermitian theory to the same form and match their coefficients, we must abandon all terms which can only be generated naturally in the $i\phi^3$ theory.

Moreover, if \eqref{eq-ziphi3t} describes the same physical system with the original $i\phi^3$ theory, both must have the same physical mass. Denote the physical mass of the $i\phi^3$ theory as $m_{\mathrm{P},i\phi^3}$ whose square is minus single-particle pole of the full propagator in momentum space
\begin{equation}\label{eq-ppiphi3}
\tilde{D}_{i\phi^3}(p)\equiv\int \frac{d^dy\ e^{-ipy}}{(2\pi)^d}\frac{1}{Z_{\mathcal{PT}}[0]}\int D\phi\exp\left\{-\int d^dx\left[\frac12\phi_x\left(-\partial^2_x+m^2\right)\phi_x+ig\phi^3_x\right]\right\}\phi_y\phi_0,
\end{equation}
 we have
\begin{equation}\label{eq-piphi3}
m^2+2\left(\frac{13}{2}+c_4\right)\frac{g^2}{m^2}D_0=m^2_{\mathrm{P},i\phi^3}.
\end{equation}
With $c_1\sim c_5$ solved, \eqref{eq-ziphi3t} simplifies to be
\begin{equation}\label{eq-ziphi3ts}\begin{aligned}
Z_{\mathcal{PT}}[j]=&\int D\phi\exp\left\{-\int d^dx\left[\frac12\phi_x\left(-\partial^2_x+m^2\right)\phi_x+\eta(\rho(\phi_x))j_x\right.\right.\\
&\left.\left.+\left(\frac{13}{2}+c_4\right)\frac{g^2}{m^2}D_0\phi^2_x+\left(-c_4-2\right)\frac{g^2}{m^2}D_0^2+3g^2\int d^dyD_{y}^3+\mathcal{O}(g^3)\right]\right\},
\end{aligned}\end{equation}
where $c_4$ is understood to be the solution of \eqref{eq-piphi3}.

As we have transformed the $\mathcal{PT}$-symmetric $i\phi^3$ theory into a free theory, we should find a Hermitian theory that can be transformed into the same form. From our previous work\cite{lyd2023}, the isospectral local Hermitian Hamiltonian of $H=\frac12p^2+\frac12m^2x^2+igx^3$ is
\begin{equation}\label{eq-preh}
h=\frac12p^2+\frac12m^2x^2+\frac{5g^2}{2m^2}x^4-\frac{g^2}{2m^4}+\mathcal{O}(g^3),
\end{equation}
and naturally we can assume the $i\phi^3$ and $\phi^4$ correspondence up to $g^2$-order holds also in $d$ dimensions. We find that it is indeed the case.

Consider the Euclidean partition function of $\phi^4$ in $d$ dimensions,
\begin{equation}\label{eq-zphi4}\begin{aligned}
Z_{\mathrm{Herm}}[j]=\int D\phi\exp\left\{-\int d^dx\left[\frac12\phi_x\left(-\partial^2_x+m^2\right)\phi_x+\lambda_1\phi^4_x+\lambda_2+i\phi_xj_x\right]\right\},
\end{aligned}\end{equation}
where $\lambda_1$ and $\lambda_2$ are assumed to be of the same order, and we will denote higher order terms as $\mathcal{O}(\lambda^2)$. Making use of the completing-the-square method again, we make the following ansatz of the transformation $\kappa(\phi_x)$ which is expected to transformation the $\phi^4$ theory into a free theory,
\begin{equation}\label{eq-transkappa}\begin{aligned}
\kappa\left(\phi_x\right)=&\phi_x+\int d^dy\left[f_1\lambda_1D_{x-y}\phi^3_y+f_2\lambda_1 D_0D_{x-y}\phi_y+\mathcal{O}(\lambda^2)\right],
\end{aligned}\end{equation}
where $f_1$ and $f_2$ are real coefficients.

Under the transformation \eqref{eq-transkappa}, $Z_{\mathrm{Herm}}[j]$ is as follows,
\begin{equation}\label{eq-zphi4t}\begin{aligned}
Z_{\mathrm{Herm}}[j]=\int D\phi\exp&\left\{-\int d^dx\left[\frac12\phi_x\left(-\partial^2_x+m^2\right)\phi_x+\left(f_2-3f_1\right)\lambda_1 D_0\phi^2_x-f_2\lambda_1 D_0^2+\lambda_2+i\kappa(\phi_x)j_x\right.\right.\\
&\left.\left.+\left(1+f_1\right)\lambda_1\phi^4_x+\mathcal{O}(\lambda^2)\right]\right\}.
\end{aligned}\end{equation}
To obtain a free theory which is also isospectral to the $\phi^4$, the following equations must be satisfied,
\begin{equation}\label{eq-pphi4}
1+f_1=0,\ m^2+2\left(f_2-3f_1\right)\lambda_1 D_0=m^2_{P,\phi^4},
\end{equation}
where $m^2_{P,\phi^4}$ is minus single-particle pole of the full propagator in momentum space
\begin{equation}\label{eq-ppphi4}
\tilde{D}_{\phi^4}(p)\equiv\int \frac{d^dy\ e^{-ipy}}{(2\pi)^d}\frac{1}{Z_{\mathrm{Herm}}[0]}\int D\phi\exp\left\{-\int d^dx\left[\frac12\phi_x\left(-\partial^2_x+m^2\right)\phi_x+\lambda_1\phi^4_x+\lambda_2\right]\right\}\phi_y\phi_0.
\end{equation}
With $f_1$ and $f_2$ solved, \eqref{eq-zphi4t} simplifies to be
\begin{equation}\label{eq-zphi4ts}\begin{aligned}
Z_{\mathrm{Herm}}[j]=\int D\phi\exp&\left\{-\int d^dx\left[\frac12\phi_x\left(-\partial^2_x+m^2\right)\phi_x+\left(f_2+3\right)\lambda_1 D_0\phi^2_x-f_2\lambda_1 D_0^2+\lambda_2+i\kappa(\phi_x)j_x+\mathcal{O}(\lambda^2)\right]\right\},
\end{aligned}\end{equation}
where $f_2$ is understood to be the solution of \eqref{eq-pphi4}.

If both \eqref{eq-ziphi3ts} and \eqref{eq-zphi4ts} describe the same physical system, we must have the following equations,
\begin{equation}\label{eq-12eta}\begin{aligned}
 \left(\frac{13}{2}+c_4\right)\frac{g^2}{m^2}=&(f_2+3)\lambda_1,\\
\left(-c_4-2\right)\frac{g^2}{m^2}D_0^2+3g^2\int d^dyD_{y}^3=&-f_2\lambda_1 D_0^2+\lambda_2,\\
\eta(\rho(\phi_x))=&\kappa(\phi_x),
\end{aligned}\end{equation}
from which $\lambda_1,\lambda_2,\eta(\phi_x)$ can be solved as follows,
\begin{equation}\label{eq-12etas}\begin{aligned}
\lambda_1=&\frac{(13+2c_4)g^2}{2(f_2+3)m^2},\\
\lambda_2=&\frac{(9f_2-6c_4-12)g^2}{2(f_2+3)m^2}D_0^2+3g^2\int d^dyD_{y}^3,\\
\eta(\phi_x)=&\kappa(\rho^{-1}(\phi_x)),
\end{aligned}\end{equation}

So far, we have work out the isospectral local Hemitian theory of the $\mathcal{PT}$-symmetric $i\phi^3$ theory up to $g^2$-order in $d$ dimensions, which is a $\phi^4$ theory given by \eqref{eq-zphi4} and \eqref{eq-12etas}. 

In $1$ dimension, the calculation of propagators is quite straightforward. From \eqref{eq-ppiphi3} and \eqref{eq-ppphi4}, $\tilde{D}_{i\phi^3}(p)$ and $\tilde{D}_{\phi^4}(p)$ are as follows,
\begin{equation}\begin{aligned}
\tilde{D}_{i\phi^3}(p)=&\frac{1}{2\pi}\frac{1}{p^2+m^2+\frac{9g^2}{m^3}+\frac{18g^2}{m(p^2+4m^2)}+\mathcal{O}(g^3)},\\
\tilde{D}_{\phi^4}(p)=&\frac{1}{2\pi}\frac{1}{p^2+m^2+\frac{6\lambda_1}{m}+\mathcal{O}(\lambda^2)},
\end{aligned}\end{equation}
from which $m^2_{\mathrm{P},i\phi^3}$ and $m^2_{\mathrm{P},\phi^4}$ is solved to be
\begin{equation}\begin{aligned}
m^2_{\mathrm{P},i\phi^3}=m^2+\frac{15g^2}{m^2}+\mathcal{O}(g^3),\ m^2_{\mathrm{P},\phi^4}=m^2+\frac{6\lambda_1}{m}+\mathcal{O}(\lambda^2).
\end{aligned}\end{equation}
Consequently, $c_4$ and $f_2$ can be solved explicitly to be $c_4=\frac{17}{2}$ and $f_2=3$ and thus $\lambda_1=\frac{5g^2}{2m^2}$ and $\lambda_2=-\frac{g^2}{2m^4}$, which give exactly the same result of the isospectral Hermitian theory as \eqref{eq-preh}. 

The matching of our method in this work and the method in our previous work\cite{lyd2023} supports strongly that our method in this work is a kind of generalization of our previous method from quantum mechanics to quantum field theory. Moreover, all equations in this section have the same form for any positive value of $d$ and differ in coefficients only, which means results in quantum mechanics are also useful in general $d$ dimensions. In our previous work\cite{lyd2023}, we start from a $\mathcal{PT}$-symmetric Hamiltonian and end with an isospectral local Hermitian Hamiltonian in a sequential manner, without any assumption of the form of the isospectral Hermitian theory. In this work, we assume the form of the isospectral local Hermitian theory and match its coefficients with the coefficients of the $\mathcal{PT}$-symmetric theory. With the help of results in our previous work, assumptions of the form of the isospectral Hermitian theory are no longer necessary because the form of the isospectral Hermitian theory in $d$ dimensions is fixed as long as that in $1$ dimension is fixed!

\section{The $i\phi^3$ in $0$ dimensions and numerical results}\label{sec-d0}

In $0$ dimensions, we solve $\rho(\phi_x)$ and $\kappa(\phi_x)$ numerically to show they exist and perturbative expansions \eqref{eq-transrho} and \eqref{eq-transkappa} work well at small $\phi_x$ values. Because $\mathcal{PT}$ symmetry is left-right symmetry in the complex plane\cite{bender-book}, it is enough to study $i\phi^3$ and $\phi^4$ on half of the real axis, and we constrain that $\phi\in[0,\phi_{\max}]$.

To transform the $i\phi^3$ in $0$ dimensions into a free theory, we should solve the following equation
\begin{equation}\label{eq-0iphi3}\begin{aligned}
\rho'(\phi)\exp\left(-\frac12m^2\rho(\phi)^2-ig\rho(\phi)^3\right)=\exp\left(-\frac12m_{\mathrm{P},i\phi^3}^2\phi^2-\Delta E_{i\phi^3}\right),
\end{aligned}\end{equation}
where $m_{\mathrm{P},i\phi^3}^2$ is given by \eqref{eq-ppiphi3} as follows,
\begin{equation}\begin{aligned}
\frac{1}{m_{\mathrm{P},i\phi^3}^2}=\frac{\int_{-\infty}^{\infty}d\phi\exp\left(-\frac12m^2\phi^2-ig\phi^3\right)\phi^2}{\int_{-\infty}^{\infty}d\phi\exp\left(-\frac12m^2\phi^2-ig\phi^3\right)},
\end{aligned}\end{equation}
and $\Delta E_{i\phi_3}$ is the extra vacuum energy,
\begin{equation}\begin{aligned}
\int_{-\infty}^{\infty}d\phi\exp\left(-\frac12m^2\phi^2-ig\phi^3\right)=\int_{-\infty}^{\infty}d\phi\exp\left(-\frac12m_{\mathrm{P},i\phi^3}^2\phi^2-\Delta E_{i\phi^3}\right).
\end{aligned}\end{equation}
To solve \eqref{eq-0iphi3} numerically, a boundary condition is necessary, which we choose to be
\begin{equation}\begin{aligned}
\exp\left.\left(-\frac12m^2\rho(\phi)^2-ig\rho(\phi)^3\right)\right|_{\phi=\phi_{\max}}=\exp\left.\left(-\frac12m_{\mathrm{P},i\phi^3}^2\phi^2-\Delta E_{i\phi^3}\right)\right|_{\phi=\phi_{\max}},
\end{aligned}\end{equation}
which demands $\rho(\phi)$ does not have exponential behavior at large $\phi$. Setting $m=1,g=0.1,\phi_{\max}=100$, the numerical result of $\rho(\phi)$ and its comparison with the perturbative result are shown in FIG. \ref{fig-1} and FIG. \ref{fig-2}. The numerical result shows $\tan^{-1}\left(\frac{\mathrm{Im}(\rho(\phi_{\max}))}{\mathrm{Re}(\rho(\phi_{\max}))}\right)=-0.16\pi$, which means $\rho(\phi)$ is in the Stokes sector whose boundaries are $\mathrm{arg}\rho=0$ and $\mathrm{arg}\rho=-\frac13\pi$ which is the correct Stokes sector defining the $i\phi^3$\cite{bender-book}. FIG. \ref{fig-2} shows that perturbation theory works well in the small $\phi$ regime.

\begin{figure}
\centering
\begin{minipage}{0.49\linewidth}
\includegraphics[width=0.9\linewidth]{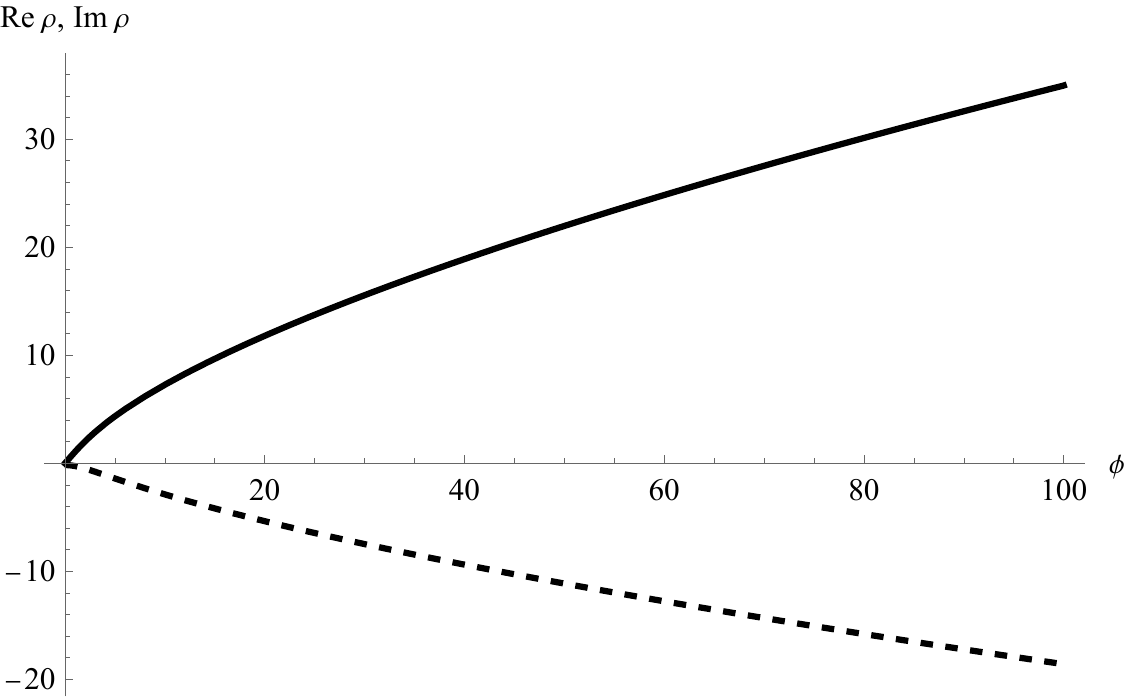}
\caption{The numerical result of $\rho(\phi)$ in $\phi\in[0,\phi_{\max}]$. $\mathrm{Re}\rho$ is the black solid line and $\mathrm{Im}\rho$ is the black dashed line.}
\label{fig-1}
\end{minipage}
\begin{minipage}{0.49\linewidth}
\includegraphics[width=0.9\linewidth]{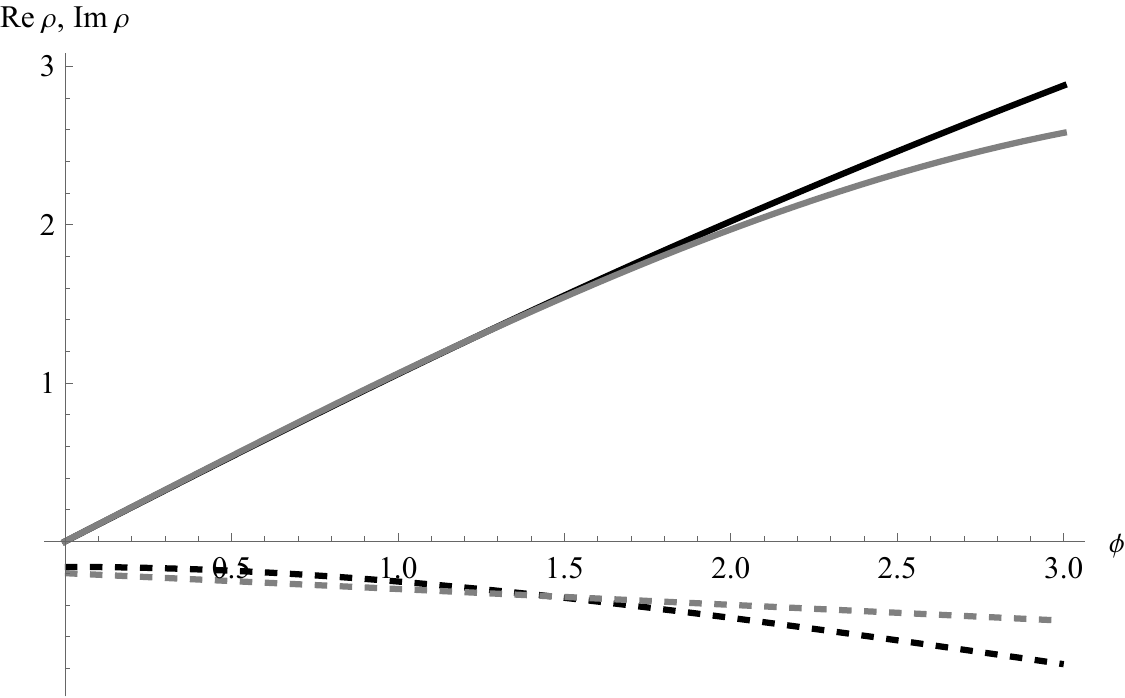}
\caption{Comparison between the numerical result and the perturbative result of $\rho(\phi)$ in $\phi\in[0,3]$. In $0$ dimensions, \eqref{eq-transrho} is $\rho(\phi)=\phi+\frac{17g^2}{2m^6}\phi-\frac{5g^2}{2m^4}\phi^3-\frac{ig}{m^2}\phi^2-\frac{2ig}{m^4}$.  For the numerical result, $\mathrm{Re}\rho$ is the black solid line and $\mathrm{Im}\rho$ is the black dashed line. For the perturbative result, $\mathrm{Re}\rho$ is the gray solid line and $\mathrm{Im}\rho$ is the gray dashed line.}
\label{fig-2}
\end{minipage}
\end{figure}

Similarly, to transform the $\phi^4$ in $0$ dimensions into a free theory, we should solve the following equation
\begin{equation}\label{eq-0phi4}\begin{aligned}
\kappa'(\phi)\exp\left(-\frac12m^2\kappa(\phi)^2-\lambda\kappa(\phi)^4\right)=\exp\left(-\frac12m_{\mathrm{P},\phi^4}^2\phi^2-\Delta E_{\phi^4}\right),
\end{aligned}\end{equation}
where $m_{\mathrm{P},\phi^4}^2$ is given by \eqref{eq-ppphi4} as follows,
\begin{equation}\begin{aligned}
\frac{1}{m_{\mathrm{P},\phi^4}^2}=\frac{\int_{-\infty}^{\infty}d\phi\exp\left(-\frac12m^2\phi^2-\lambda\phi^4\right)\phi^2}{\int_{-\infty}^{\infty}d\phi\exp\left(-\frac12m^2\phi^2-\lambda\phi^4\right)},
\end{aligned}\end{equation}
and $\Delta E_{\phi^4}$ is the extra vacuum energy,
\begin{equation}\begin{aligned}
\int_{-\infty}^{\infty}d\phi\exp\left(-\frac12m^2\phi^2-\lambda\phi^4\right)=\int_{-\infty}^{\infty}d\phi\exp\left(-\frac12m_{\mathrm{P},\phi^4}^2\phi^2-\Delta E_{\phi^4}\right).
\end{aligned}\end{equation}
To solve \eqref{eq-0phi4} numerically, a boundary condition is necessary, which we choose to be
\begin{equation}\begin{aligned}
\exp\left.\left(-\frac12m^2\kappa(\phi)^2-\lambda\kappa(\phi)^4\right)\right|_{\phi=\phi_{\max}}=\exp\left.\left(-\frac12m_{\mathrm{P},\phi^4}^2\phi^2-\Delta E_{\phi^3}\right)\right|_{\phi=\phi_{\max}},
\end{aligned}\end{equation}
which demands $\kappa(\phi)$ does not have exponential behavior at large $\phi$. Setting\footnote{From \eqref{eq-12etas}, $\lambda_1=3\frac{g^2}{m^2}$ in $0$ dimensions, so we set $\lambda=0.03$ to ensure the $i\phi^3$ and the $\phi^4$ are evaluated at approximately the same interaction strength.} $m=1,\lambda=0.03,\phi_{\max}=100$, the numerical result of $\kappa(\phi)$ and its comparison with the perturbative result are shown in FIG. \ref{fig-3} and FIG. \ref{fig-4}. FIG. \ref{fig-4} shows again that perturbation theory works well in the small $\phi$ regime.

\begin{figure}
\centering
\begin{minipage}{0.49\linewidth}
\includegraphics[width=0.9\linewidth]{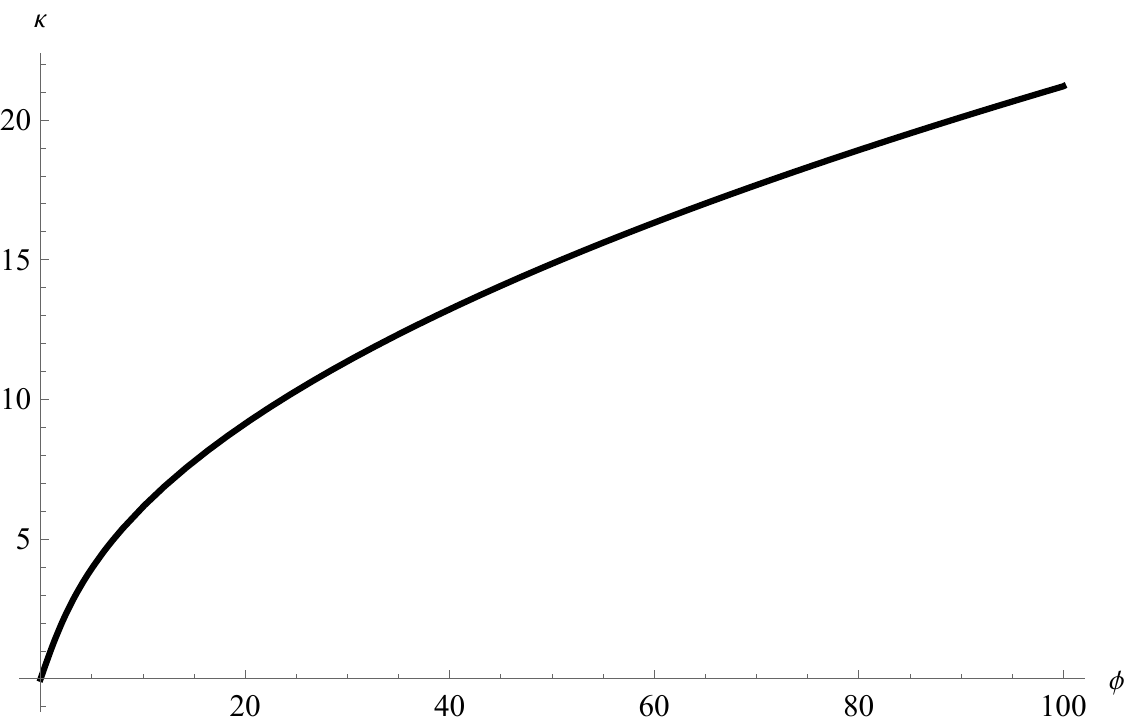}
\caption{The numerical result of $\kappa(\phi)$ in $\phi\in[0,\phi_{\max}]$.}
\label{fig-3}
\end{minipage}
\begin{minipage}{0.49\linewidth}
\includegraphics[width=0.9\linewidth]{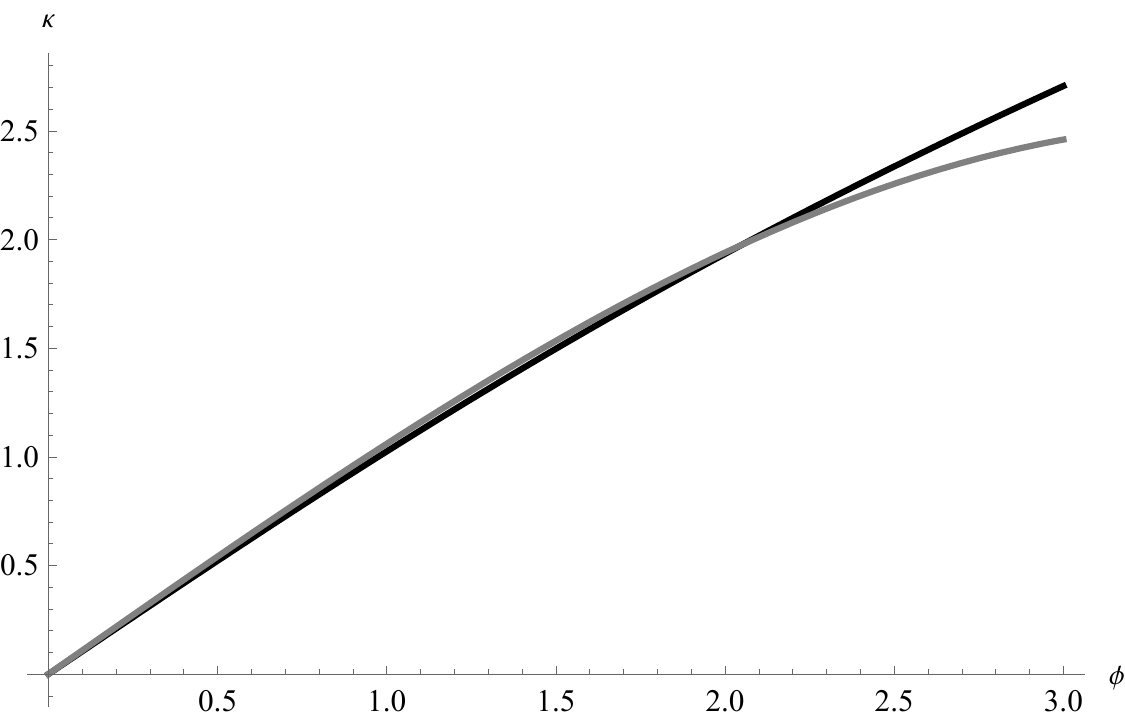}
\caption{Comparison between the numerical result and the perturbative result of $\kappa(\phi)$ in $\phi\in[0,3]$. In $0$ dimensions, \eqref{eq-transkappa} is $\kappa(\phi)=\phi-\frac{\lambda}{m^2}\phi^3+\frac{3\lambda}{m^4}\phi$. For the numerical result, $\kappa$ is the black solid line. For the perturbative result, $\kappa$ is the gray solid line.}
\label{fig-4}
\end{minipage}
\end{figure}

\section{Summary and outlook}\label{sec-sum}

In this work, we calculate the isospectral local Hermitian theory of the $\mathcal{PT}$-symmetric $i\phi^3$ quantum field theory at $g^2$-order in $d$ dimensions, which turns out to be a $\phi^4$ theory, making use of a method which is a generalization of the method in our previous work\cite{lyd2023}. The main idea is that we transform both the $\mathcal{PT}$-symmetric theory and the assumed Hermitian theory into the same form, and then match their coefficients. We work in $d$ dimensions and find that the derivation is similar in all dimensions, which means our results in quantum mechanics\cite{lyd2023} can be generalized directly to $d$ dimensions using the framework build in this work.

In our approach, we find that matching the single-particle pole is enough to determine the coefficients of the transformation in \eqref{eq-transrho} from the $i\phi^3$ theory to its isospectral local Hermitian theory at $g^2$-order. It is reasonable to raise the question that whether the multi-particle contribution such as the spectral function in the K\"all\'{e}n-Lehmann is necessary or not in the matching procedure, especially at higher order. Actually, the physical field $\phi_{\mathrm{phys},x}$ is determined only after we determine the transformation from the $i\phi^3$ theory to its isospectral local Hermitian theory, and it is not possible to calculate the physical spectral function of the $i\phi^3$ theory before the matching procedure. However, as we calculate the transformation only up to $g^2$-order, there may be additional terms at higher order in the ``free'' form after the $\phi_x\rightarrow\rho(\phi_x)$ transformation, which do not depend on the form of the original interaction and represent multi-particle contribution, and we left the discussion of higher order terms to future works.  At $g^2$-order, considering only the single-particle pole in the matching procedure is exact, and the coefficients in \eqref{eq-transrho} is neither under-constrained nor over-constrained.

As far as we know, there are few results about the isospectral local Hermitian theory of the $\mathcal{PT}$-symmetric $i\phi^3$ quantum field theory, which is usually believed to be nonlocal\cite{bender-book}, and the locality of our result is fresh in the study of the $\mathcal{PT}$-symmetric quantum theory. The locality of the isospectral local Hermitian theory may help us to reveal the physical meaning of a $\mathcal{PT}$-symmetric theory, and may simplify a lot in the calculation of physical observables such as scattering amplitudes. We believe many further works can be conducted based on this work in future.

\end{widetext}

\bibliography{ref.bib}

\end{document}